\documentstyle[floats,aps]{revtex}
\def\BRA{\left\langle}
\def\KET{\right\rangle}
\def\LK{\left(}
\def\RK{\right)}
\def\LBK{\left\lbrack}
\def\RBK{\right\rbrack}
\def\LB{\left\lbrace}
\def\RB{\right\rbrace}
\begin{document}
\draft
\twocolumn[\hsize\textwidth\columnwidth\hsize\csname
@twocolumnfalse\endcsname

\title{Localization Length Exponent,
Critical Conductance Distribution and Multifractality
in Hierarchical Network Models for the Quantum Hall Effect}
\author{Andreas Weymer, Martin Janssen\\\
Institut f\"ur Theoretische Physik, Universit\"at zu
K\"oln,\\\  Z\"ulpicher Strasse 77, 50937 K\"oln, Germany}
\date{May 05, 1998}
\maketitle

\begin{abstract} 
We study hierarchical network models which have recently been
introduced to approximate the Chalker-Coddington model for the integer
quantum Hall effect (A.G. Galstyan and M.E. Raikh, PRB {\bf 56} 1422
(1997); Arovas et al., PRB {\bf 56}, 4751 (1997)).  The hierarchical
structure is due to a recursive method starting from a finite
elementary cell.  The localization-delocalization transition occurring
in these models is displayed in the flow of the conductance
distribution under increasing system size. We numerically determine
this flow, calculate the critical conductance distribution, the
critical exponent of the localization length, and the multifractal
exponents of critical eigenstates. 
\end{abstract}

\vskip2pc]

\bigskip

\section{Introduction}\label{secint}
Localization brought about by quantum interference in disordered media
\cite{And58} is an ongoing research topic. One reason for this lies in
the complex structure of the associated localization-delocalization
(LD) transition, a quantum phase transition which has no simple mean
field description \cite{JanR98}.  Although many aspects of the LD
transition are understood and quantitative results for critical
exponents have been obtained by extensive numerical studies (see
e.g. \cite{Kra93}) there is no complete theory available yet. Best
understood are quasi-one-dimensional (1D) systems, where the notion of
a transfer-matrix led to strong analytic methods (see
e.g. \cite{Mir94,BeeR97}).  Non-perturbative renormalization
techniques have also been developed on the basis of quasi-1D
geometries. However, no such technique proved to be successful
directly for higher dimensional systems.  There have been early
attempts in that direction \cite{Lee,Aok} based on real-space
renormalization within a Hamiltonian description of the systems
dynamics. A major obstacle against their success was the increase in
the number of relevant matrix elements of the effective Hamiltonian
after a few renormalization-group (RG) steps --- unless the system was
in a regime of strong localization.  Only recently two groups have
independently introduced a novel real-space renormalization technique
for the LD problem \cite{Gal97,Aro97}.  It was realized that a local
scattering theoretical formulation (see e.g. \cite{Sha82}) of the
systems dynamics gives rise to a straightforward real-space RG which
is not restricted to quasi-1D.  In particular, they considered the
Chalker-Coddington (CC) model \cite{Cha88}, a two-dimensional network
of nodes and links where waves can propagate on the links and are
scattered by chiral scatterers placed on the nodes.  This model is
very convenient for testing the RG ideas since the scatterers are described
by simple $2\times 2$ random $S$-matrices.  It has been established
\cite{Cha88,Kiv93,Kle95} that the CC model contains interesting LD
physics.  The chiral structure is meant to capture the essential
features of the motion of 2D electrons in a strong perpendicular
magnetic field and a random potential \cite{Cha88}.  Such electron
systems exhibit the integer quantum Hall effect and are denoted as
quantum Hall systems \cite{PraGirv}.  The quantum Hall effect is
essentially caused by a LD transition \cite{JanB}. In a localized
state the electron is confined to some finite area within the bulk of
the system while the electron's wave function is extended over the
entire system in a delocalized state. However, states carrying a
unidirectional quantized edge-current exist also in the localized
phase of a quantum Hall system.  The link amplitudes of the CC model
can be identified with local wave functions of a quantum Hall system.
On increasing the average scattering strength $\BRA T\KET $ from $0$
to $1$ one finds localization and chiral edge states (carrying a
quantized Hall current) in almost any case. However, as soon as $\BRA
T\KET $ is close to $0.5$ extended bulk states appear, and e.g. the
Hall conductance jumps to another quantized plateau value.  It is
expected that, in the thermodynamic limit, the transition occurs at
precisely $\BRA T\KET =T^{\ast}=0.5 $. 
At the plateau to plateau transition the localization length $\xi$ 
of a critical
state diverges, $\xi \propto |\BRA T\KET -T^{\ast}|^{-\nu}$, where the critical
exponent  $\nu=2.35\pm 0.03$ \cite{Huc92}. Furthermore, it is known that the
states are multifractal as soon as their localization length becomes
larger than system size \cite{Kle95}.

It turned out, that the qualitative aspects of the LD transition could
be described by a simple Migdal-Kadanoff type approach \cite{Aro97}
leading to simple algebraic RG equations. Critical exponents, however,
were found to be  different from the commonly accepted
values. Quantitatively more reliable results were obtained by
\cite{Gal97} when using a slightly more realistic approximation to the
CC model. In that case, however, the treatment needed some numerical
assistance. Despite the success in describing the LD transition the RG
approach was based on uncontrolled approximations.

It was therefore tempting to elaborate further on the RG ideas of
\cite{Gal97,Aro97} and develop a general scheme which enables to
improve quantitative results in a systematic manner.  Advantageous for
that purpose is the interpretation of the RG in terms of hierarchical
networks embedded in the CC model.  The hierarchical models are
constructed from elementary cells cut out of the CC model and are
characterized by the number of nodes, $b^2$, of the elementary cell
and a number of active scatterers, $V$.  For each model the flow of
the conductance distribution, the critical conductance distribution
and the critical exponent of the localization length can be determined
by iterative RG steps.  Furthermore, a calculation of multifractal
properties of critical states in the hierarchical models allows to
complete the description of LD transition characteristics. It turns
out that our RG approach leads to reasonable values already for
moderate sizes of elementary cells.

We have organized the paper as follows.  In Sec.~2 we introduce the
hierarchical models and describe their connection to those considered
in \cite{Gal97,Aro97}.  In Sec.~3 the flow of the conductance
distribution is analyzed.  The critical exponent of the localization
length and the critical conductance distributions are determined. The
multifractal analysis of critical states in hierarchical models forms
the content of Sec.~4, followed by our conclusions in Sec.~5.

\section{Hierarchical Network Models}\label{sechir}
We recall the construction of the Chalker-Coddington (CC) network
 model.  A set of elementary scatterers is put on the nodes of a
 regular 2D lattice. Unidirectional channels link the scatterers to
 each other as shown in Fig.~1.  The elementary scatterers are
 described by a unitary $2\times 2$ scattering matrix $S$,
\begin{equation}
\left( \begin{array}{c} O \\ O' \end{array} \right) = 	S
\left( \begin{array}{c} I \\ I' \end{array} \right)\; ,\;\;
	S = \left( \begin{array}{cc} t & r' \\ r & t'  \end{array}
		\right)\, .\label{2.1}
\end{equation}
$T=|t|^2=|t'|^2$ is the transmission to the left and $R=|r|^2=|r'|^2
=1-T$ is the transmission to the right.  The phases on the links are
taken as random variables uniformly distributed between $0$ and
$2\pi$. Also the transmission (`scattering strength') $T\in [0,1]$ can
be taken from a certain distribution.

To study the localization-delocalization behavior of the CC network
model one can impose boundary conditions for a square network of
linear size $L$ such that only total transmission to the left $T(L)$
or right $R(L)=1-T(L)$ is possible.  In an appropriate
two-probe conductance measurement (see Fig.~2) the transmission $T(L)$
yields the conductance $G$ via the Landauer-B\"uttiker theory of
coherent transport,
\begin{equation}
	G=\frac{e^2}{h}T\, .
\end{equation}
 Thus, an interesting object to study the LD transition is the
distribution function ${ P}_L(T)$ and its flow with increasing $L$.
In the thermodynamic limit, $L\to \infty$, bulk localized phases are
characterized by $T=0$ or $T=1$. In both cases current can only flow
along edge states. Thus, two stable (attractive) fixed point
distributions ${ P}_{\rm I} (T)=\delta(T)$ and ${ P}_{\rm II}
(T)=\delta(T-1)$ exist.  The LD transition will be characterized by a
third fixed point distribution, denoted as ${ P}^{\ast}(T)$, which is
{unstable} (repulsive). The unstable character means that ${
P}^{\ast}$ is the limiting distribution, $\lim_{L\to \infty} { P}_L
(T)={ P}^{\ast}(T)$, only if the set of its {\em relevant} parameters
$X_i(L)$ (e.g. average value, median, variance)
is initially taken to the fixed point values $X_i^{\ast}$.
In any other case the distribution approaches one of the two stable
fixed point distributions, $P_{\rm I,II}$.  For large enough initial
sizes $L_0$ it may turn out that one parameter $X(L_0)$ is enough to
fix the flow of the whole distribution. This situation is referred to
as one-parameter-scaling  regime. The corresponding {
$\beta$-function} $\beta(X)=d X/d\ln L$ describes the scaling flow of
$X(L)$ and allows for calculating the critical exponent $\nu$ of the
correlation length (localization length)\cite{NOTE},
\begin{equation}
	\xi\propto |X -X^{\ast}|^{-\nu}\, ,\label{2.2}
\end{equation}
 from 
\begin{equation}
 \nu^{-1} = \frac{d \beta}{d X} \Bigg|_{X^{\ast}}
	\,. \label{2.3}
\end{equation} 

Therefore, a possible way of determining the critical exponent $\nu$
would be to set up a renormalization-group (RG) process for the flow
of ${P}_L(T)$. An exact process could be achieved by recording the
values $T(L)$ for arbitrary system sizes $L$ and for an arbitrary
large number of realizations of the random scatterers. Such process
could be carried out with the help of numerical calculations and is
only limited by the computational capacity.  In a more efficient
way, however, one replaces the network model by some {\em
hierarchically constructed} network that allows to perform the
renormalization recursively.  The hierarchical structure requires some
finite elementary cell which allows for calculating the corresponding
transmission $T$ analytically or by simple numeric routines. However,
the replacement of the original network by a hierarchical one
introduces some error which is not under control.  To reach definite
answers for critical exponents one has to repeat the RG process for
larger and larger elementary cells such that eventually the original
model can be recovered. This procedure is advantageous if (i) one is
aiming at qualitative results and (ii) the results for the critical
exponents rapidly converge.  To get qualitative results about the LD
transition the elementary cell should be as simple as possible, still
containing essentials like basic symmetries. For example, the chiral
structure and paths with loops in the CC network model should be
maintained in order to allow for quantum interference along oriented
paths. This idea was used in \cite{Aro97} where a simple elementary
cell with $4$ active scatterers (those for which neither $T=0$ nor
$T=1$) was considered (see Fig.~6 in \cite{Aro97}). In that
approximation the scattering turned out to obey one-dimensional
composition laws which helped to proceed analytically.  Unfortunately,
in this approximation a discrete mirror symmetry (with respect to the
central vertical axis) is broken.  As a result, the critical
distribution ${ P}^\ast(T)$ is not symmetric with respect to $T=0.5$,
and the critical average transmission is different from $0.5$.

The obvious choice for an elementary cell respecting also the mirror
symmetry is shown in Fig.~3a. It consists of a $3\times3$ lattice
containing $5$ active scatterers and $4$ scatterers (in the corners)
that fully transmit to the left ($T=1$) or to the right ($T=0$).
Building a hierarchical network from such elementary cells leads to a
fractal network with fractal dimension $D^{[3/5]}=\ln 5/\ln 3\approx
1.46$ (see Fig.~3a).  It will be denoted as $\LBK 3/5\RBK$-model.  The
corresponding RG process has been already carried out in
Ref.~\cite{Gal97}. In that work a different interpretation was
adopted. As in the $\LBK 3/5\RBK$-model an elementary cell of $5$
scattering units is replaced by one in each renormalization
step(see Fig.~2 in \cite{Gal97}).
 As shown in Fig.~3b the scaling factor is $2$ instead
of $3$. In other words, the RG process of \cite{Gal97} proceeds by
constructing a new network from renormalized scatterers while our
process considers a hierarchical network as a fractal subset of the
original CC network.  Apart from the different interpretation both
processes are, of course, equivalent if the difference in scaling
factors is taken into account.  The simplified elementary cell,
considered in Fig.~6 of \cite{Aro97} corresponds to the $\LBK
3/5\RBK$-network when replacing the central scatterer by a fully
reflecting/transmitting one (see Fig.3 c) and will henceforth be
denoted as $\LBK 3/4\RBK$-model. Its fractal dimension is
$D^{[3/4]}=\ln 4/\ln 3\approx 1.26$.

In the present work we extend the RG process to larger elementary
cells and study also the multifractal properties of critical
wave functions.

A generalization to larger elementary cells is straightforward:
each Chalker-Coddington network  with $ b\times b$ sites
($b=5,7,9 \ldots$) 
and boundary conditions of only 4 links to the outer world 
 can serve as an elementary cell. 
By recursion one gets a
 hierarchical infinite network (see Fig.~3a,c).
Its fractal dimension is 
\begin{equation}
	D^{[b/V]}= \frac{\ln V}{\ln b} \, \label{2.4}
\end{equation}
where the elementary cell contains $V=(b^2+1)/2$ scatterers.
Note, that for $b\to \infty$ the original CC model is recovered and e.g.
 the fractal dimension approaches $2$ for $b\gg 1$.

In general, hierarchical networks can be characterized by an odd
number $b$ describing the size of its elementary cell ($b^2$) and the
number $V$ of active scatterers.  Accordingly we denote the
hierarchical networks as $\LBK b/V\RBK$-network.

In addition to the hierarchical networks with $V(b)=(b^2+1)/2$ which,
 for large $b$, reduce to the original CC model one can use $\LBK
 b/V\RBK$-elementary cells to construct larger elementary cells of
 size $b'\times b'$ with $V(b')< (b'^2+1)/2$ active scatterers, in
 order to simplify the RG steps.  We elaborated on a $\LBK
 9/37\RBK$-model constructed from $\LBK 3/5\RBK$-elementary cells
 (displayed in Fig.~8). Note that the number of active scatterers
 ($V(9)=37$) is less then $(9^2+1)/2=41$ which would follow from the
 straightforward construction for $b=9$.
\section{Flow of the Conductance Distribution}\label{secrgd}
After having outlined which hierarchical models will be considered in
this work we are now going to describe how the RG for the conductance
distribution was obtained in our  numerical calculations. 

1. In the first step we algebraically solve the scattering problem for
a particular elementary cell. We obtain the transmission
probability $T(L=b)$ as a function of the scattering matrix elements
$t=|t|e^{i\alpha}$, $t'=|t|e^{i\alpha'}$, $r=|r|e^{i\beta}$ and
$r'=|r|e^{i\beta'}$ of all active scatterers. Here the unitarity
constraints $$e^{i(\alpha +\alpha')}=-e^{i(\beta +\beta')}$$ and
$$|t|^2+|r|^2=1$$ have to be obeyed.  That leaves for $T(b)$ an
algebraic expression of $V$ scattering strengths, $T_{1,\ldots, V}$,
and $3V$ phases.

2. In the second step we draw the scattering strengths $T$ from an
initial distribution, $P_0(T)$ peaked around a certain mean value
$\BRA T\KET_0$ and random uncorrelated phases for $\alpha,\alpha'$ and
$\beta$.  For each realization we calculated the scattering strength
$T(L=b)$ according to the algebraic expression.

3. After collecting a large number ($\approx 2000$) of scattering
strengths we calculated the corresponding histogram and normalized it.
That yields an approximation for the distribution function
$P_{L=b}(T)$ depending on the initial distribution $P_0(T)$.  We also
determined the average value $\BRA T\KET (L=b)$ and other quantities
like median or geometric mean of the distribution.

4. We repeat the steps No. 2 and 3, but now drawing $T$ from the
previously obtained distribution $P_{L=b}(T)$.  We did this for each
active scatterer independently.  As a result we got a novel
distribution which should coincide with $P_{L=b^2}(T)$ of the
hierarchical $\LBK b/V\RBK$-model if the phases at each RG step can be
taken as uncorrelated random phases. Indeed, the phases of
$t(L=b),r(L=b),t'(L=b)$ turned out to be  uniformly
distributed in the interval $ [0,2\pi[ $ and correlations were
not observed. Therefore,  the 4th step yields a
reasonable approximation for $P_{L=b^2}(T)$ and, in particular, for
$\BRA T\KET (L=b^2)$.

5. Next, we repeat step No. 4 $N$ times ($N\approx 20$) which yields,
for each initial distribution $P_0(T)$, the flow of the distribution
function $P_{L}(T)$ and of the mean $\BRA T\KET (L)$ as functions of
$\ln L=(N+2)\ln b$.

6. We repeated the RG process for a number of different initial
distributions $P_0(T)$.

An example for the flow of the distribution is displayed in Fig.~4 for
a $\LBK 3/4\RBK$-system with initial value $\BRA T\KET < T_c\approx
0.6$. It demonstrates that the distribution is attracted by the
localization fixed point at $T=0$.

It generally turned out that after a few iterations the
mean $\BRA T\KET$ was already a good scaling variable, i.e. once a
certain mean value was reached, the flow of the distribution after
further iterations did no longer depend significantly on the initial
distribution.  In particular we found for each model a certain
distribution which was independent of $L$, characterized by $\BRA
T\KET (L) \equiv T^\ast$ but unstable against changes $\BRA T\KET
-T^\ast\not=0$.  Thus, we took the mean $\BRA T\KET $ to investigate
the instability of the critical distribution.  The critical exponent
$\nu$ of the localization length is determined  as
\begin{equation}
	\nu = \frac{(N-N_0)\ln b}{\ln \LB (T(N)- T^\ast)/(T(N_0)-T^\ast)\RB}
	\, .\label{2.5}
\end{equation}
This expression follows from Eq.~(\ref{2.3}) with $X= \BRA T\KET$. The
numbers $N_0$ and $N$ were chosen such that $\BRA T\KET$ was
significantly repelled from $T^\ast$ without leaving the regime of
power law repulsion.
 
In  Figs.~5 -- 8 the elementary cells together with 
 the critical distribution are shown. For the $[3/5]$, $[9/37]$-models also 
the flow of the average transmission is displayed.   
 The critical distribution turned out to be broadly distributed over
 the interval $\LBK 0,1\RBK$.  Those models respecting the mirror
 symmetry of its elementary cell showed critical distributions
 approximately symmetric with respect to $0.5\approx T^\ast$.  The
 $\LBK 3/4\RBK$-model breaks the mirror symmetry and $T^\ast\approx
 0.6$, as previously found in \cite{Aro97}.  This model and the $\LBK
 3/5\RBK$-model discussed in \cite{Gal97} are taken as a reference to
 judge the improvement in quantitative results when going to larger
 elementary cells. The symmetric critical distributions for different
 models are hardly distinguishable when taking numerical uncertainties
 into account.  The critical distribution is not 
 uniform over the interval $\LBK 0,1\RBK$ (as conjectured in
 \cite{Kog96}), but develops a unique shallow minimum at $0.5$ and has
 maximum values at the extreme positions $T=0,1$. This confirms the
 observations of \cite{Gal97} for the $\LBK 3/5\RBK$-model and of
 \cite{Fis96Lee96} for the original CC model.
 
The values for the critical exponent $\nu$ are collected in
Tab.~1. One observes that with increasing elementary cells the
critical exponent $\nu$ approaches the CC network value in a monotonic
way from above and the $\LBK 9/37\RBK$ value, $\nu= 2.75\pm 0.3$, is
already close to the value $\nu= 2.43\pm 0.18$ obtained for the
original CC model \cite{Kiv93}.  Note that the value $\nu=3.5\pm 0.3$
for the $\LBK 3/5\RBK$-model is compatible with the result
($\nu\approx 2.4$) of \cite{Gal97} when taking the difference in
scaling factors into account ($\ln 3/\ln 2\approx 1.58$).
\section{Multifractal Analysis of Critical States}\label{secmul}
A rather complete characterization of a LD transition phenomenon in
terms of critical exponents is provided by the knowledge of the
critical exponent $\nu$ of the localization length and the
multifractal exponent $\alpha_0$ of critical states \cite{JanR98}.
The exponent $\alpha_0$ describes how the typical value $P_{\rm t}$ of
the probability $P$ to find an electron (being in that state) in a
small box of linear size $l$ scales with respect to $\lambda=l/L$,
\begin{equation}
	P_{\rm t} (\lambda) \propto \lambda^{\alpha_0} \, . \label{3.1}
\end{equation} 
The fact that critical states have $\alpha_0> D$ in a $D$-dimensional
system shows that critical states are extended but not totally `space
filling' (multifractal). Indeed, the typical value scales faster to
zero as the average value ($ \BRA P \KET \propto L^{-D}$) and thus
their ratio
\begin{equation}
	\hat{\rho}:=P_{\rm t}/\BRA P\KET \propto
	L^{-(\alpha_0-D)}\label{3.11} 
\end{equation}
 is a convenient order-parameter for the LD transition. It vanishes
algebraically on approaching the LD transition point from the extended
phase and stays zero in the localized phase \cite{JanR98}. Also in the
quantum Hall system, where no extended phase exists, $\alpha_0$
essentially determines the scaling of critical states, including
energetic and spatial correlations.

The multifractal exponent $\alpha_0$ has been numerically determined
in \cite{Kle95} for the original CC model and found to be $2.27\pm
0.01$. The method used in that work can be described briefly as
follows: The network model with its scattering matrices gives rise to
a discrete unitary time evolution operator $U$ operating on bond
amplitudes and mapping a state $\psi(t)$ at some time $t$ to a novel
amplitude configuration $\psi(t+1)$ at one instant of time later by
taking all of the scattering conditions into account.  Energy
eigenstates $\psi$ are solutions of
\begin{equation}
	U\psi =\psi \label{3.71}
\end{equation}
and thus correspond to the eigenstates of $U$ with eigenvalue $1$.
The energy of the underlying electron system enters here only
parametrically through the scattering matrices and $U$ will have an
eigenstate to eigenvalue $1$ only for a discrete set of the parameter
$T$.  Instead of looking for this discrete set one can fix $T$ and ask
for the more general eigenvalue problem
\begin{equation}
     U(T)\psi_n(T)= e^{i\phi_n(T)} \psi_n(T) \, .\label{3.72}
\end{equation} 
$\psi_n(T)$ are, for fixed $T$, a set of eigenstates at scattering strength
$T$ of slightly modified disorder configurations, each of the
modifications being an overall shift in random phases on the
links. 
 
The typical probability $P_{\rm t}(\lambda)$ can be identified as the
geometric mean of local squared amplitudes integrated over boxes of
size $l$,
\begin{equation}
	P_{\rm t} (\lambda):= \exp \BRA \ln \LK \sum_{\rm box}
	|\psi|^2\RK\KET \, . \label{3.3}
\end{equation}  
Here the average $\BRA \ldots \KET$ can be performed over different
realizations and/or different local boxes within one realization.

We adopted this method for  computing the critical exponent $\alpha_0$
of the hierarchical network models. The scattering strengths were
fixed to the critical value $T^\ast$.  We could not rely on a
recursive calculation method, but needed some finite system of linear
size $L>b$ where $b$ is the linear size of the elementary cell. This
was necessary to detect the strong fluctuations of local amplitudes
responsible for the multifractal behavior. Instead of going to very
large sizes $L$ we took moderate sizes ($L \approx 50$) but used
several realizations to obtain $\alpha_0$ via
Eqs.~(\ref{3.3},\ref{3.1}). An example for a critical wave function of
a $\LBK 3/5\RBK$-system is shown in Fig.~9.
Strong spatial correlations,
 previously
observed in critical states \cite{Pra96}, 
are even more pronounced on the fractal support.
 
Since the
hierarchical networks are defined on geometric fractals of dimension
$D^{[b/V]}$ it is inappropriate to compare different values of
$\alpha_0$ directly. Rather one should address the exponent of the
order-parameter $\hat{\rho}$ which is $\alpha_0-D^{[b/V]}$. These
values are listed in Tab.~1. It can be seen that $\alpha_0-D^{[b/V]}$
monotonically approaches the CC model value from above as $D^{[b/V]}$
approaches $d=2$ from below.  Since large values for $\alpha_0-D$
correspond to strong multifractal fluctuations of critical amplitudes
this behavior tells that the hierarchical models with sparsely
distributed active scatterers show more pronounced fluctuations than
those with a denser set of active scatterers.  For the $\LBK
9/37\RBK$-model, however, we obtain $\alpha_0-D=0.33\pm 0.05$ which is
already close to the CC value, $0.27 \pm 0.01$.
\section{Conclusions}
We extended the concept of a real-space renormalization-group approach
to the Chalker-Coddington model \cite{Gal97,Aro97} in a systematic
manner. We constructed hierarchical network models based on elementary
cells which were cut out of a Chalker-Coddington model. Each
elementary cell is characterized by the number $b^2$ of nodes and the
number $V$ of active scatterers. The hierarchical models are fractal
objects of dimension $D^{[b/V]}=\ln V/\ln b$ embedded in the original
Chalker-Coddington model.  In the $V=\propto b^2 \to \infty$ limit the
models coincide with the Chalker-Coddington  model.  We performed
renormalization-group calculations for the following $\LBK
b/V\RBK$-models: $\LBK 3/4\RBK$, $\LBK 3/5\RBK$, $\LBK 5/13\RBK $ and
$\LBK 9/37\RBK $. From these calculations we obtained the critical
conductance distribution and the critical exponent $\nu$ of the
localization length. The results for the $\LBK 3/4\RBK$ and $\LBK
3/5\RBK$-models are consistent with those obtained in
\cite{Gal97,Aro97}. Our results for larger elementary cells show how
fast the critical exponent approaches the value of the original
Chalker-Coddington model. 

Let us briefly comment on the possibility
to use our method for a precise determination of the critical
exponent of  the original Chalker-Coddington
model.  For that purpose one not only has to use larger elementary cells,
but also  has to increase the number of realizations to get smoother
histograms. Since the average value of $T$
turned out to be a good scaling variable the precision of the critical
parameter for each hierarchical model can be  improved by
increasing  the number of
realizations in each iteration step.  The critical parameters
are then more precise for the particular hierarchical model, however have
uncontrolled deviations from that of the Chalker-Coddington 
model. Better approximations
can only be gained by increasing the size of the elementary cell and by
studying the  convergence of critical exponents.  The
computational effort then drastically increases; it is expected to be
 comparable to that of the well known finite size
scaling method  based on  the transfer matrix technique \cite{Kra93}.

To complete the characterization of the
localization-delocalization transitions in the hierarchical $\LBK
b/V\RBK$-models we calculated the multifractal exponent
$\alpha_0-D^{[b/V]}$ of critical states and found that it approaches
the Chalker-Coddington value from above when going to larger
elementary cells.  In summary, our work shows that a systematic
renormalization-group treatment of networks of random quantum
scatterers can not only reveal qualitative aspects of a possible
localization-delocalization transition, but also provides quantitative
results for critical exponents the accuracy of which is controlled by
convergence under stepwise increase of the size of elementary cells.

\bigskip
\bigskip
This work was supported by the Sonderforschungsbereich 341
of the Deutsche Forschungsgemeinschaft.
We thank Boris Shapiro for stimulating us to this work and Daniel
P. Arovas for helpful discussions. MJ thanks J\'anos Kert\'esz for
drawing his attention to compositions of elementary cells.

\bigskip
\bigskip
\bigskip

\centerline{\large FIGURE CAPTIONS}

\bigskip
\bigskip

\medskip\noindent Figure 1: The graphical representation of the Chalker
Coddington network model. Wave amplitudes propagating on links can be
scattered either to the left or to the right by unitary scattering
matrices situated at the nodes of the network.

\medskip\noindent Figure 2: Schematic  representation of a
Chalker-Coddington network of size $L\times L$ in a two-probe
conductance measurement. The current flows from the source contact
(down left) through the network to the drain contact (up right).  The
conductance is given as $G(L)=(e^2/h)T(L)$.

\medskip\noindent Figure 3: a) The construction of the
$[3/5]$-hierarchical network. A $5\times 5$ cell is cut out of the
Chalker-Coddington model.  Reducing the number of links to the outer
world down to $4$ links leads to an elementary cell with $5$ active
scatterers. The hierarchical construction yields a fractal network
as a subset of the Chalker-Coddington model. At each level of the
hierarchy the system size increases by a factor of $3$. b) The
renormalization scheme of Ref.~\cite{Gal97}: 5 scatterers are replaced by
one. In each step of the renormalization the rescaling factor is $2$.
c) The hierarchical model used in Ref.~\cite{Aro97}. It can be viewed as 
 a subset of the  $[3/5]$-model with  fully transmitting/reflecting
central scatterers. 

\medskip\noindent Figure 4: Flow of the distribution of transmission
strengths under renormalization (in arbitrary units). The initial distribution
(top) is peaked around a mean-value below  the critical value $T^\ast$.
After two renormalization steps (middle \& bottom)
the distribution clearly approaches the stable distribution of full reflection
($T=0$).

\medskip\noindent Figure 5: The elementary cell of the $[3/4]$-model
 (top) and the corresponding critical distribution of scattering
 strengths (bottom).

\medskip\noindent Figure 6: The elementary cell of the $[3/5]$-model
(top left), the corresponding critical distribution of scattering
strengths (top right) and the corresponding flow of average
transmission strengths $T$ under increasing number ($N$) of 
renormalization steps (bottom).

\medskip\noindent Figure 7: The elementary cell of the $[5/13]$-model
 (top) and the corresponding critical distribution of scattering
 strengths (bottom).

\medskip\noindent Figure 8: The elementary cell of the $[9/37]$-model
(top left), the corresponding critical distribution of scattering
strengths (top right) and the corresponding flow of average
transmission strengths under increasing number ($N$) of 
renormalization steps (bottom).

\medskip\noindent Figure 9: Squared amplitudes of a critical wave function for the
$[3/5]$-model and a system size of $L=50$.

\bigskip
\bigskip
\bigskip

\centerline{\large TABLE CAPTIONS}

\bigskip
\bigskip

\noindent Table 1:  Table of models $[b/V]$, fractal dimensions
 $D$, 
 critical
exponents of the localization length $\nu$ and of the order-parameter
 exponents
$\alpha_0-D$.

\begin{references}
\bibitem{NOTE} Any smooth function $F(X)$ will  serve equally
well as a scaling variable $X'=F(X)$ leaving the critical exponent $\nu$
universal.
\end{references}
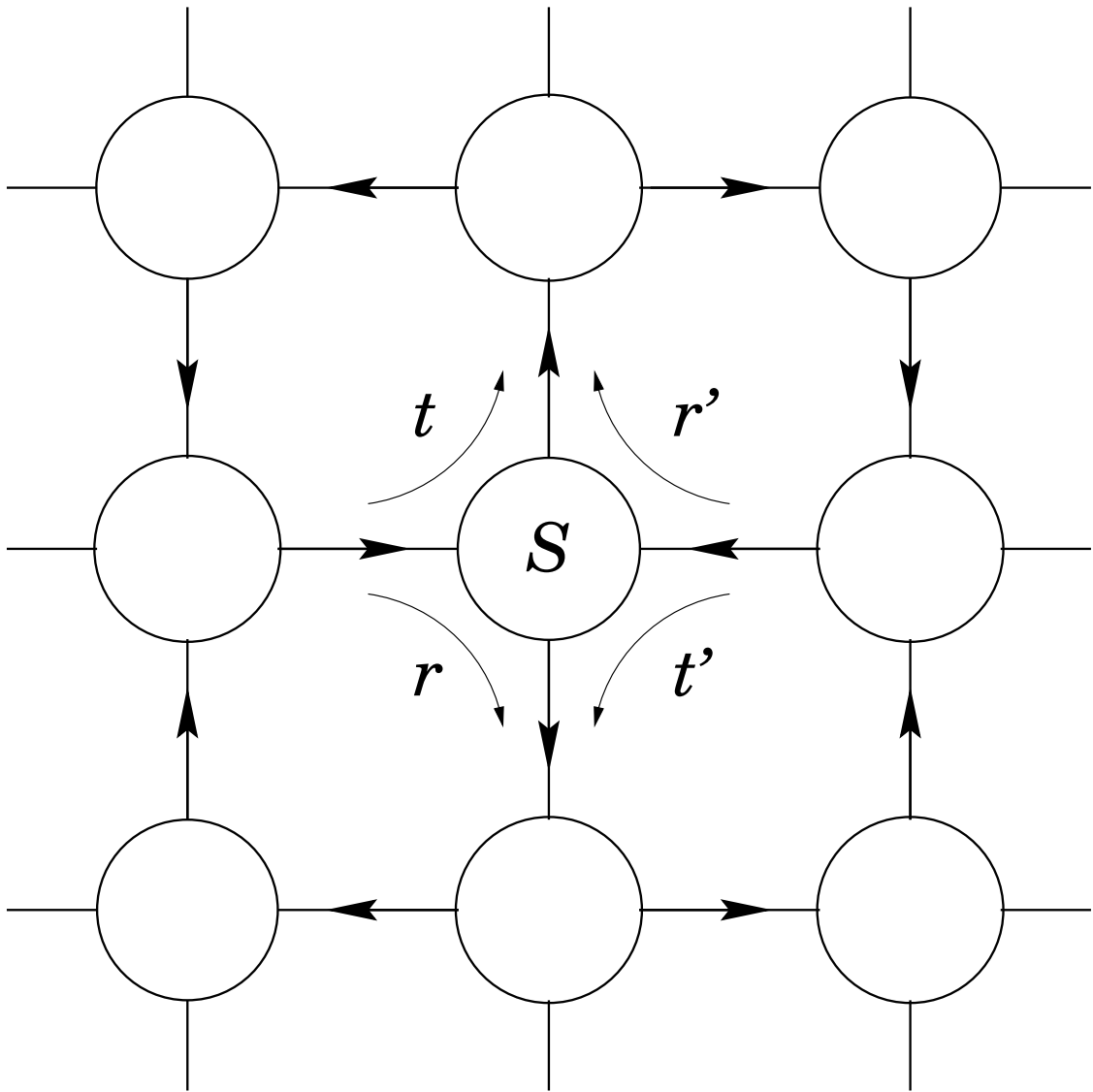
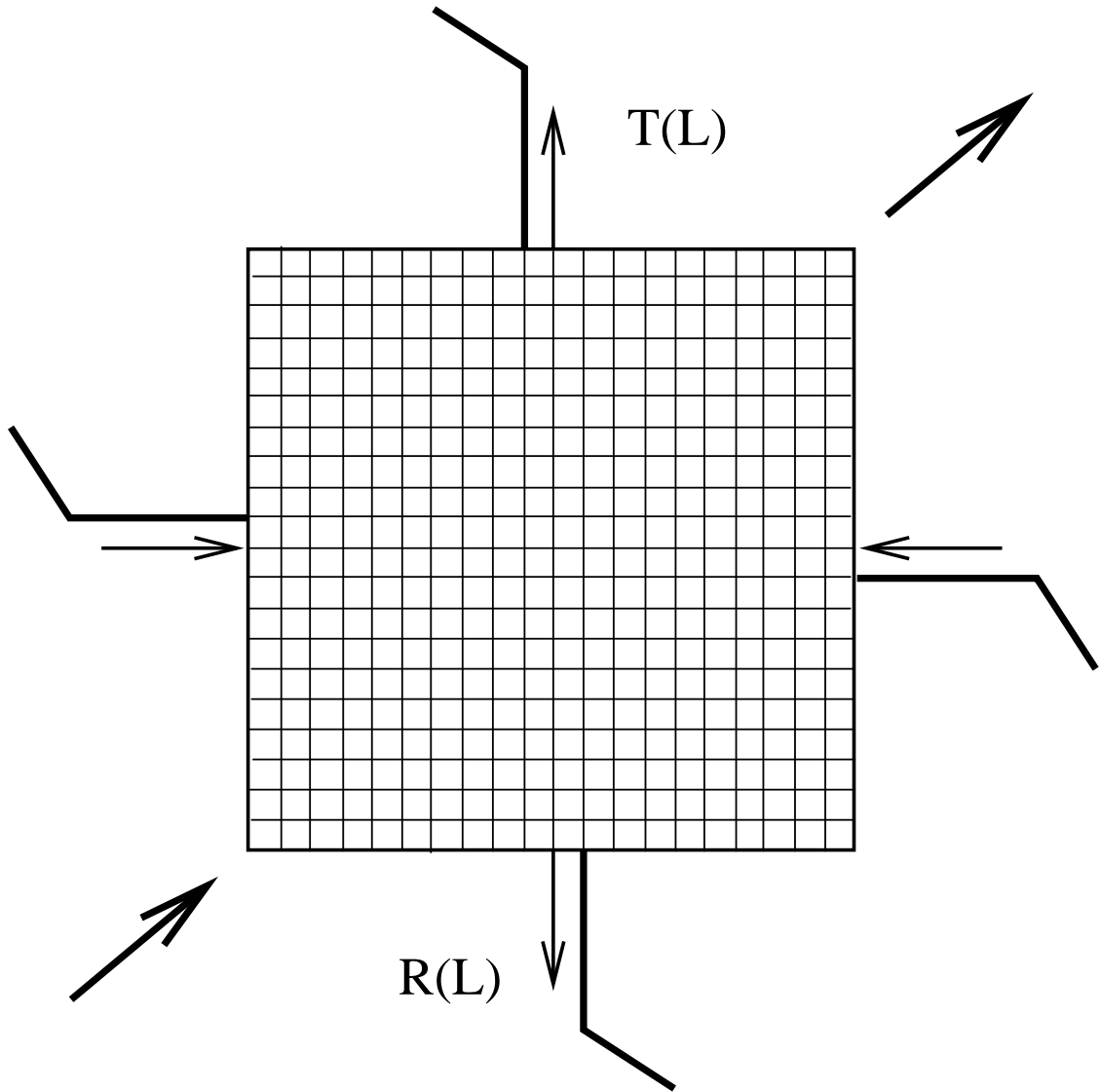
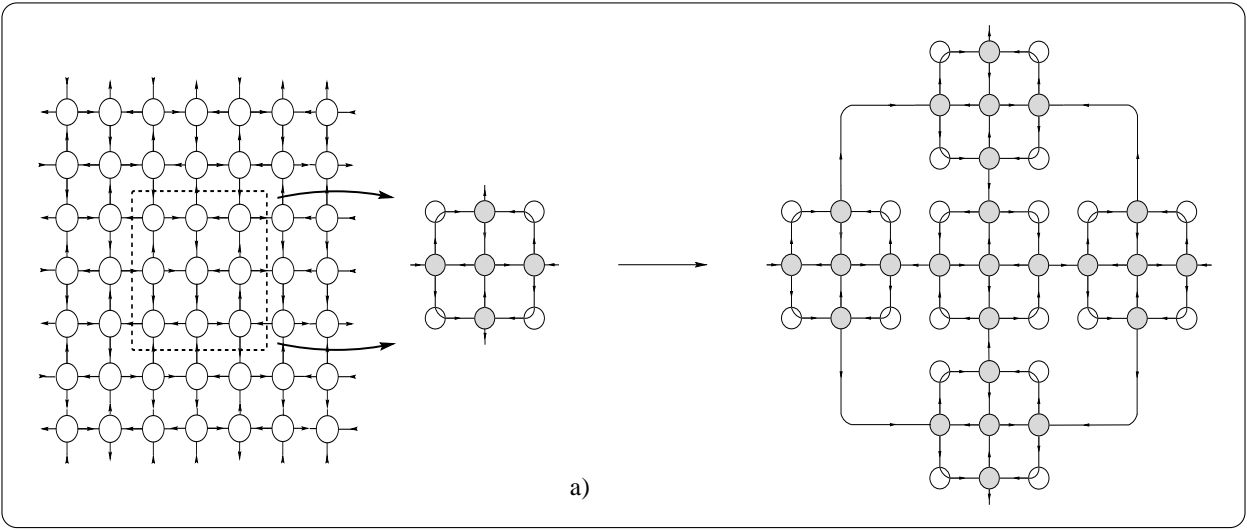
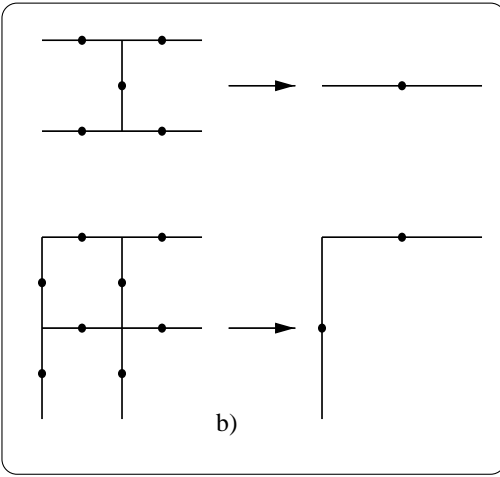
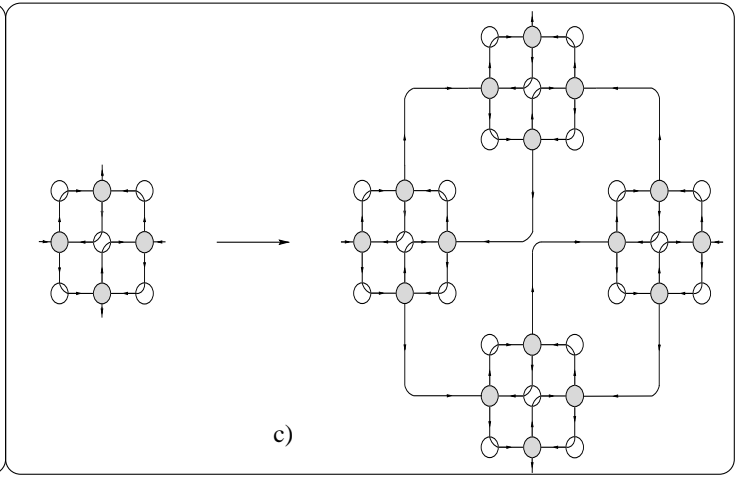
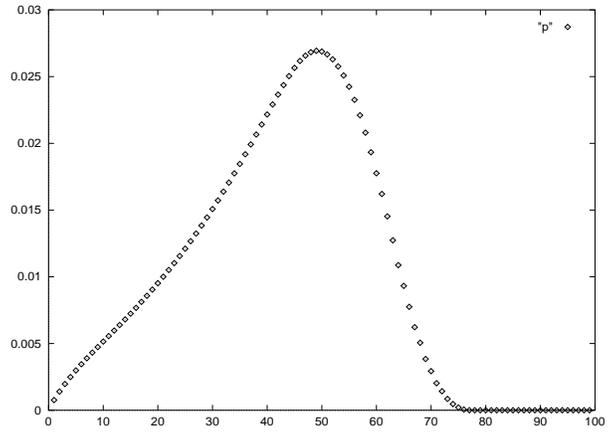
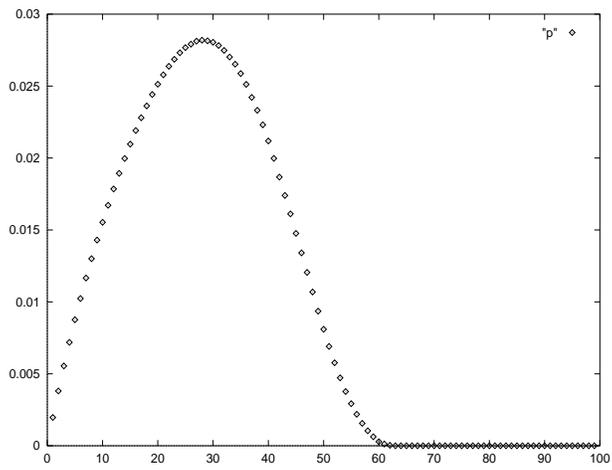
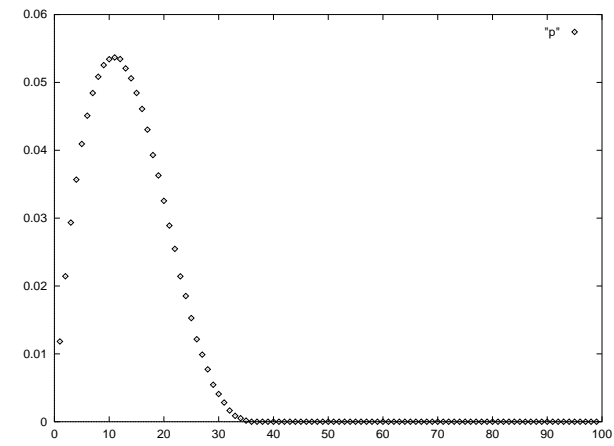
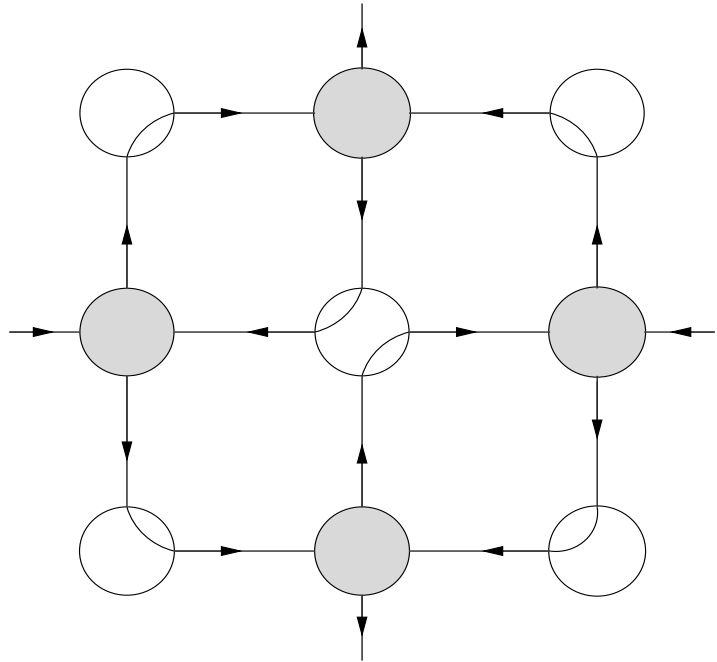
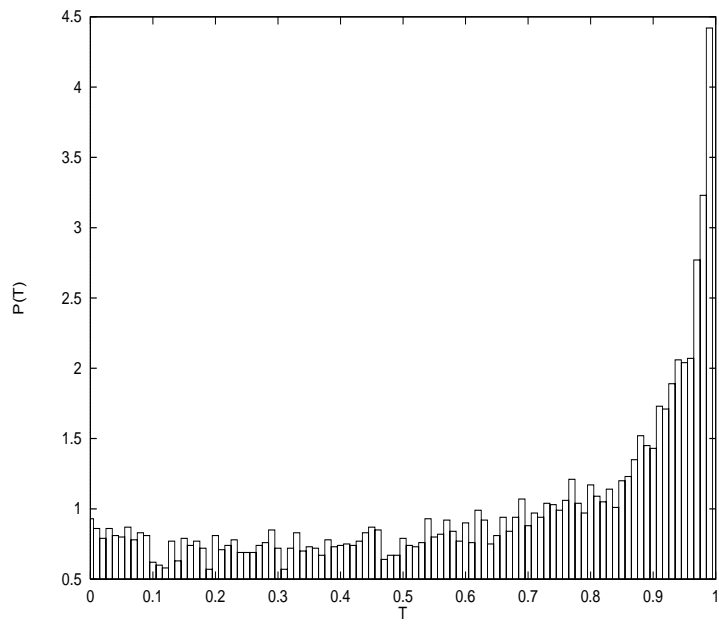
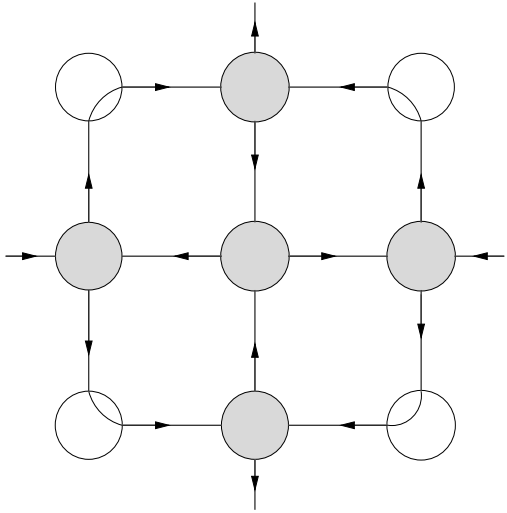
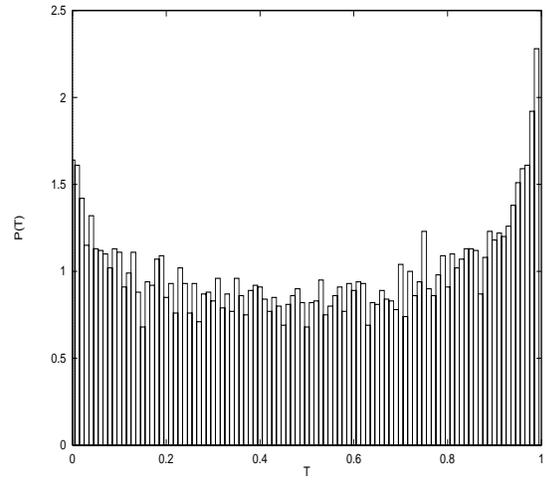
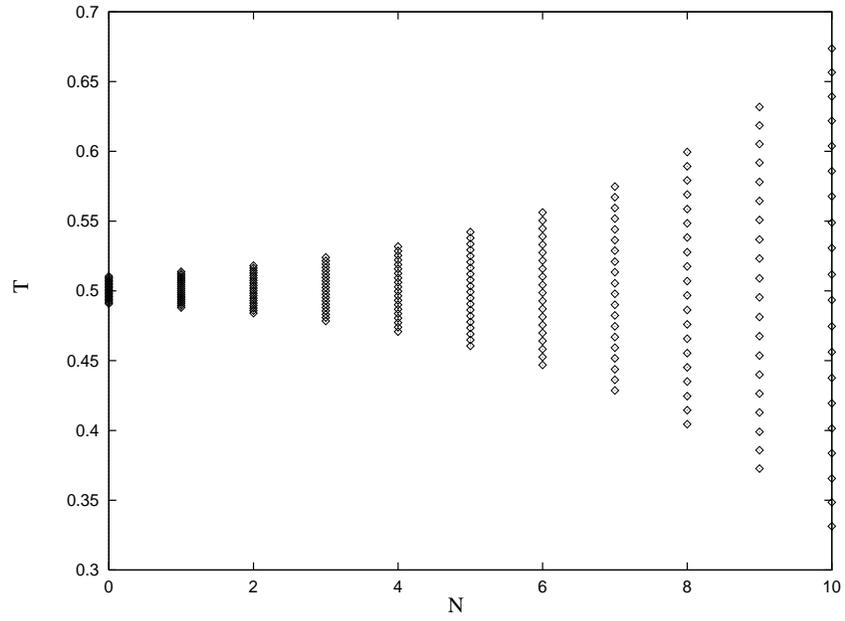
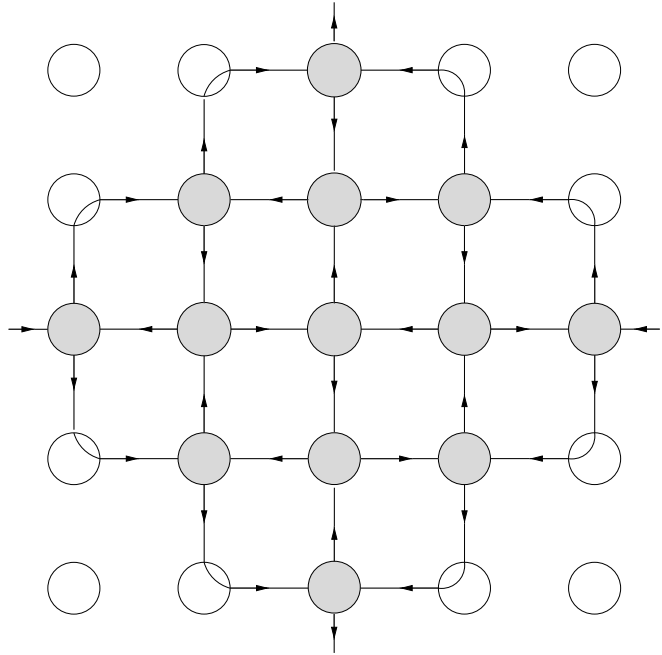
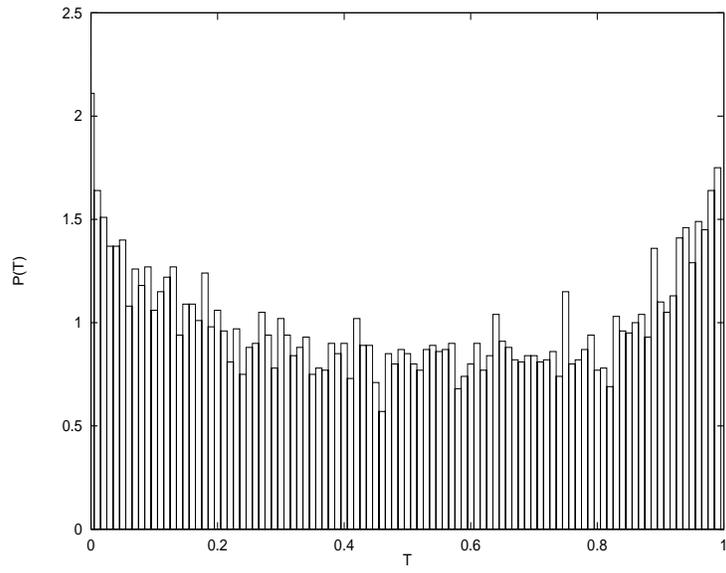
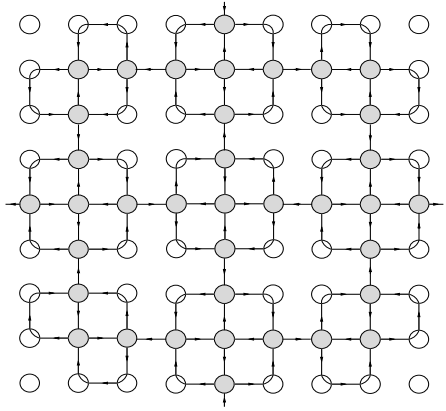
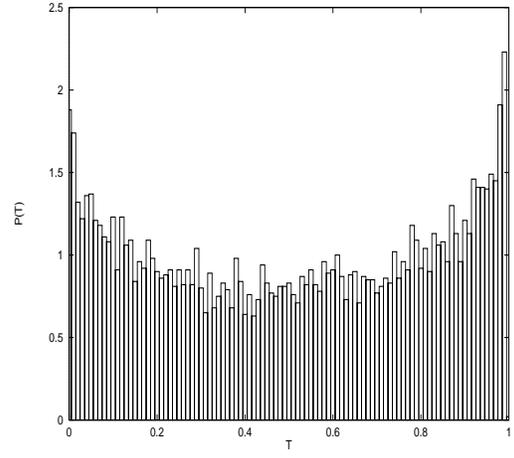
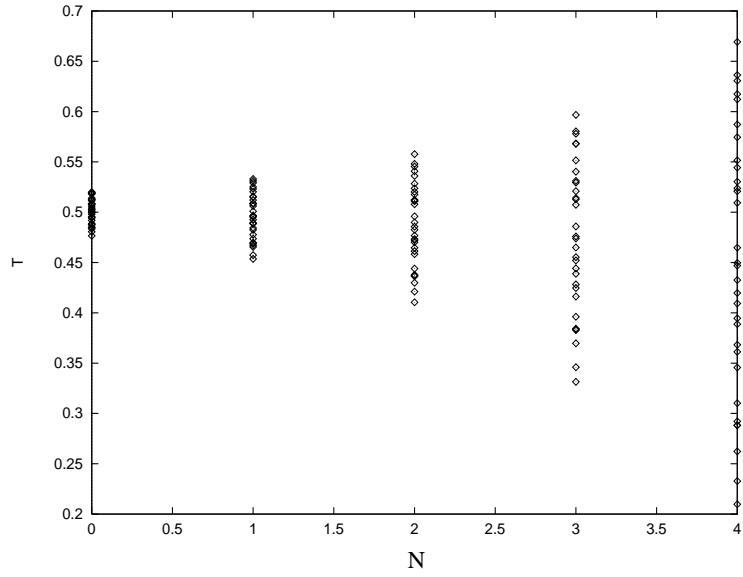
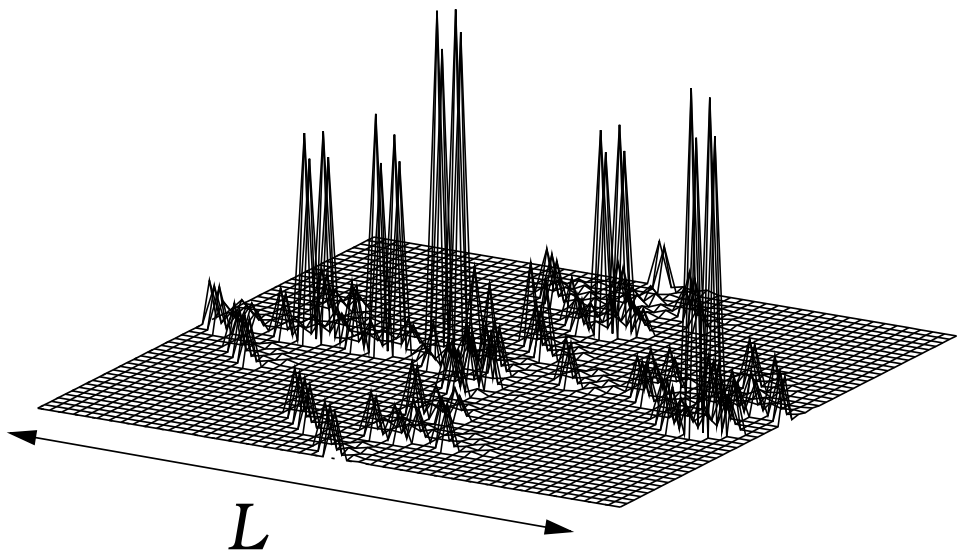
\end{document}